\documentclass{article}
\usepackage{spconf,amsmath,graphicx}
\usepackage[utf8]{inputenc} % Swedish letters
\usepackage{float}

\title{Feeding the zombies: Synthesizing brain volumes \\using a 3D progressive growing GAN}
\name{Anders Eklund$^{123}$}

\address{$^{1}$ Division of Medical Informatics, Department of Biomedical Engineering\\
    $^{2}$ Division of Statistics and Machine learning, Department of Computer and Information Science\\
    $^{3}$ Center for Medical Image Science and Visualization\\
    Linköping University, Linköping, Sweden}
\begin{document}
%\ninept
%
\maketitle

\begin{abstract}
Deep learning requires large datasets for training (convolutional) networks with millions of parameters. In neuroimaging, there are few open datasets with more than 100 subjects, which makes it difficult to, for example, train a classifier to discriminate controls from diseased persons. Generative adversarial networks (GANs) can be used to synthesize data, but virtually all research is focused on 2D images. In medical imaging, and especially in neuroimaging, most datasets are 3D or 4D. Here we therefore present preliminary results showing that a 3D progressive growing GAN can be used to synthesize MR brain volumes.
\end{abstract}

\section{Introduction}
\label{sec:intro}

Generative adversarial networks (GANs)~\cite{goodfellow2014generative} can today produce very realistic synthetic images~\cite{karras2017progressive}. GANs use adversarial training, where a generator creates an image from noise and a discriminator classifies each image as synthetic or real. During the training, the generator becomes better at generating realistic images, and the discriminator becomes better at discriminating images as synthetic or real. GANs can broadly be divided into noise-to-image GANs~\cite{goodfellow2014generative,karras2017progressive}, which produce an image from a noise vector, and image-to-image GANs, which produce an image from another image (image to image translation)~\cite{isola2017image,zhu2017unpaired}.

GANs can be particularly useful in the medical imaging field~\cite{yi2019generative}, where large datasets (e.g. data from more than 100, 1,000 or 10,000 subjects) are uncommon~\cite{bowles2018gan}. However, medical imaging data are normally not 2D, but 3D or 4D, and therefore a 3D GAN should be used to synthesize realistic volumes. Compared to 2D GANs, the current progress on 3D noise-to-image GANs is rather limited, with a few exceptions \cite{wu2016learning,wang2017shape}.

3D CycleGAN~\cite{zhu2017unpaired} has been used for image-to-image translation of medical volumes~\cite{nie2017medical,nappi2019cycle,yu20183d,abramian2019generating} and 2D noise-to-image GANs have been used to synthesize medical images~\cite{chuquicusma2018fool,bowles2018gan,frid2018gan,beers2018high,han2018gan}, but the work on 3D noise-to-image GANs in the medical imaging domain is very limited. ~\cite{mondal2018few} used a 3D GAN to synthesize MR patches of 32 x 32 x 32 voxels, but did not generate full size volumes.~\cite{kwon2019generation} used a 3D GAN to synthesize full size volumes of 64 x 64 x 64 voxels. It is difficult to judge the image quality due to the small images, but the volumes appear to lack detail. 
 
In this work, we evaluate if a 3D progressive growing GAN~\cite{karras2017progressive} can be used to synthesize realistic T1-weighted MR volumes. 

\vspace{-0.4cm}
\section{Data}
\label{sec:pagestyle}
\vspace{-0.2cm}

We used T1-weighted MR volumes from the Human Connectome Project (HCP) for training our 3D GAN\footnote{Data collection and sharing for this project was provided by the Human Connectome Project (U01-MH93765) (HCP; Principal Investigators: Bruce Rosen, M.D., Ph.D., Arthur W. Toga, Ph.D., Van J.Weeden, MD). HCP fund- ing was provided by the National Institute of Dental and Craniofacial Research (NIDCR), the National Institute of Mental Health (NIMH), and the National Institute of Neurological Disorders and Stroke (NINDS). HCP data are disseminated by the Laboratory of Neuro Imaging at the University of Southern California.} \cite{van2013wu,glasser2013minimal}. Out of the 1,113 subjects we used the first 900 subjects for training, and reserved the remaining 213 subjects for evaluations. The T1-weighted volumes were acquired with a $0.7 \times 0.7 \times 0.7$ mm isotropic voxel size, giving volumes of 260 x 311 x 260 voxels. We used preprocessed~\cite{glasser2013minimal} structural images, which have been distortion corrected and registered to MNI space. Each volume was downsampled a factor 2 and then cropped to obtain volumes of 128 x 128 x 128 voxels.

\section{Methods}

\subsection{Data augmentation}

To increase the number of training volumes we performed data augmentation by applying 10 random 3D rotations to each of the 900 volumes, to achieve a total of 9000 training volumes. The random rotations were generated from a normal distribution with a mean of 0 and a standard deviation of 10 degrees.

\vspace{-0.25cm}
\subsection{3D PGAN}

Noise-to-image GANs are rather difficult to train, especially for high resolutions, and a common pitfall is the mode collapse problem, where the GAN synthesizes a single image. A progressive growing GAN (PGAN)~\cite{karras2017progressive} is trained progressively, i.e. by first generating images of 4 x 4 pixels, then images of 8 x 8 pixels, all the way to the desired resolution. Here we investigate if the same idea can be used to synthesize high resolution volumes, by starting the training on 4 x 4 x 4 volumes, and then continue the training on 8 x 8 x 8 volumes etc.

We based our 3D PGAN on the 2D PGAN Tensorflow implementation provided in~\cite{karras2017progressive}. We replaced all 2D convolutions with 3D convolutions, and added an extra dimension to all relevant Tensorflow calls. We also added reading and writing of nifti files using the nibabel Python package~\cite{brett2016nibabel}. Our 3D PGAN is available on Github\footnote{https://github.com/wanderine/ProgressiveGAN3D} to facilitate replication of our results~\cite{eklund2017reply}. Compared to the original 2D PGAN implementation, we lowered the number of filters from 512 to 128. Table~\ref{table:parameters} states the used parameters for each resolution level, in general the learning rate is lower compared to the original 2D PGAN. The number of real volumes to show before doubling the resolution was increased from 600,000 to 1,000,000.

\begin{table}[htb]
\scriptsize
\caption{Hyperparameters used for our 3D progressive growing GAN.}
\begin{center}
\begin{tabular}{|c|c|c|}
\hline  
\textbf{\normalsize Resolution}  & \textbf{\normalsize Minibatch size}  & \textbf{\normalsize Learning rate}   \\[0.2ex]
\hline 
\normalsize  4 x 4 x 4 & \normalsize 512 & \normalsize 0.0003  \\
\normalsize  8 x 8 x 8 & \normalsize 256 & \normalsize 0.0003  \\
\normalsize  16 x 16 x 16 & \normalsize 16 & \normalsize 0.0006  \\
\normalsize  32 x 32 x 32 & \normalsize 4 & \normalsize 0.0006  \\
\normalsize  64 x 64 x 64 & \normalsize 4 & \normalsize 0.0007  \\
\hline
\end{tabular}
\end{center}
\label{table:parameters}
\end{table}

Table~\ref{table:parameters} states the training times for the 3D PGAN, using an Nvidia DGX station (containing 4 Nvidia Tesla V100 graphics cards with 32 GB of memory each). 

% run 66 to 4946
% run 70 to 6992
% run 72 to 7904
% run 78 to  

\begin{table}[htb]
\scriptsize
\caption{Approximate training times for our 3D progressive growing GAN. Each time represents the total training time to get to that resolution.}
\begin{center}
\begin{tabular}{|c|c|}
\hline  
\textbf{\normalsize Resolution}  & \textbf{\normalsize Training time}    \\[0.2ex]
\hline 
\normalsize  4 x 4 x 4 & \normalsize 8 min  \\
\normalsize  8 x 8 x 8 & \normalsize   35 min  \\
\normalsize  16 x 16 x 16 & \normalsize  4 h  \\
\normalsize  32 x 32 x 32 & \normalsize 23 h  \\
\normalsize  64 x 64 x 64 & \normalsize  81 h   \\
\hline
\end{tabular}
\end{center}
\label{table:parameters}
\end{table}

\vspace{-0.5cm}
\section{Results}
\label{sec:typestyle}

\begin{figure*}
\begin{minipage}[b]{1.0\linewidth}
  \centering
  \centerline{\includegraphics[width=1.05\textwidth]{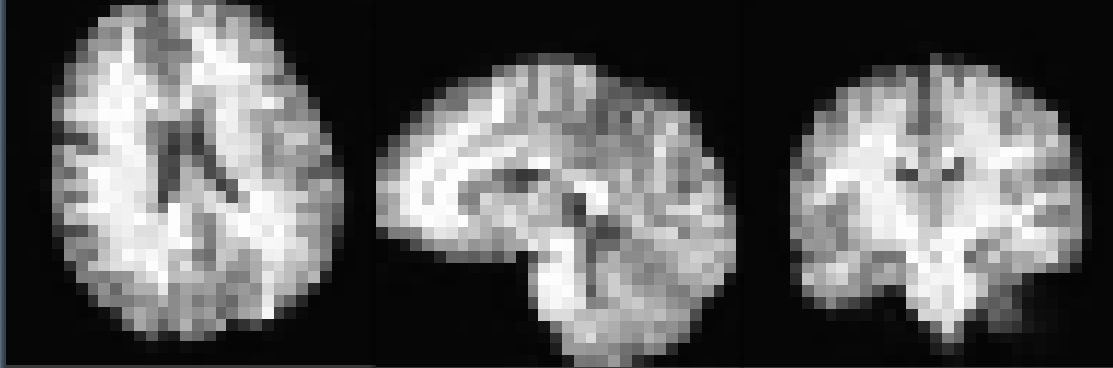}}
  \centerline{\includegraphics[width=1.05\textwidth]{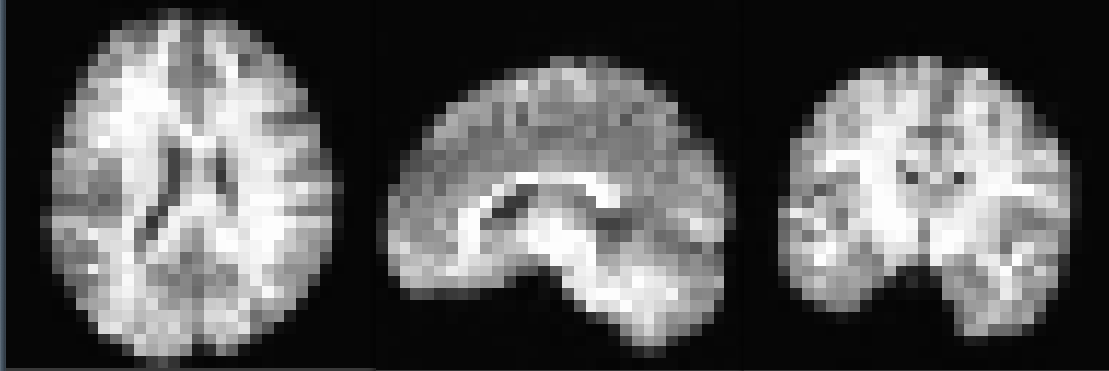}}
  \centerline{\includegraphics[width=1.05\textwidth]{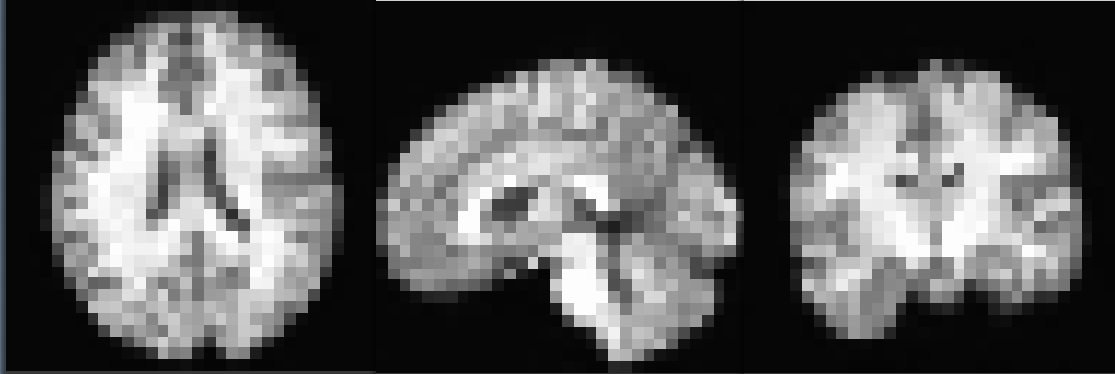}}  
\end{minipage}
\caption{Synthetic T1-weighted MR volumes with a resolution of 32 x 32 x 32 voxels (upsampled to 128 x 128 x 128 voxels) generated by our 3D progressive growing GAN.}
\label{figure1}
\end{figure*}

\begin{figure*}
\begin{minipage}[b]{1.0\linewidth}
  \centering
  \centerline{\includegraphics[width=1.05\textwidth]{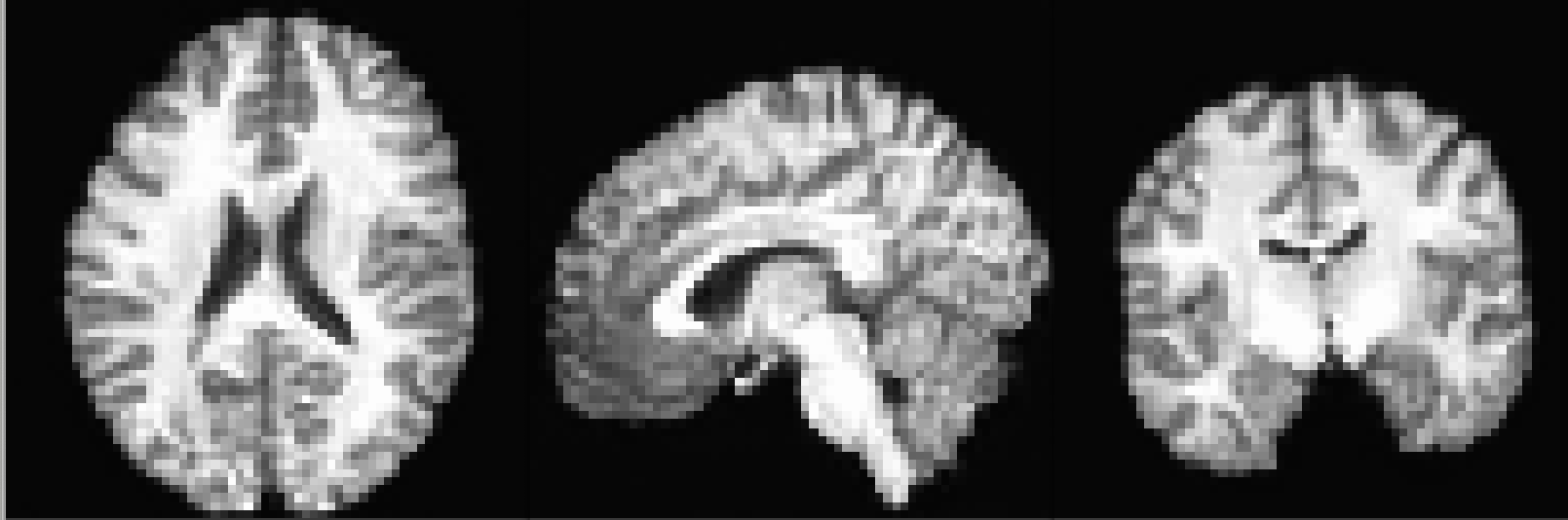}}
  \centerline{\includegraphics[width=1.05\textwidth]{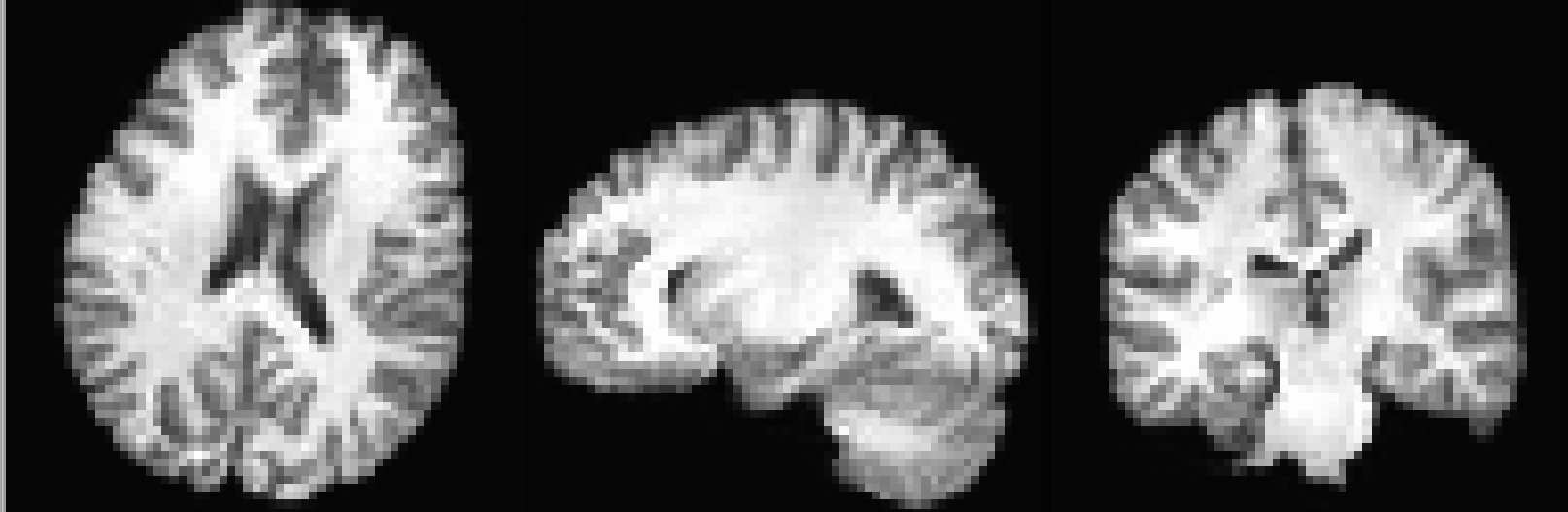}}
  \centerline{\includegraphics[width=1.05\textwidth]{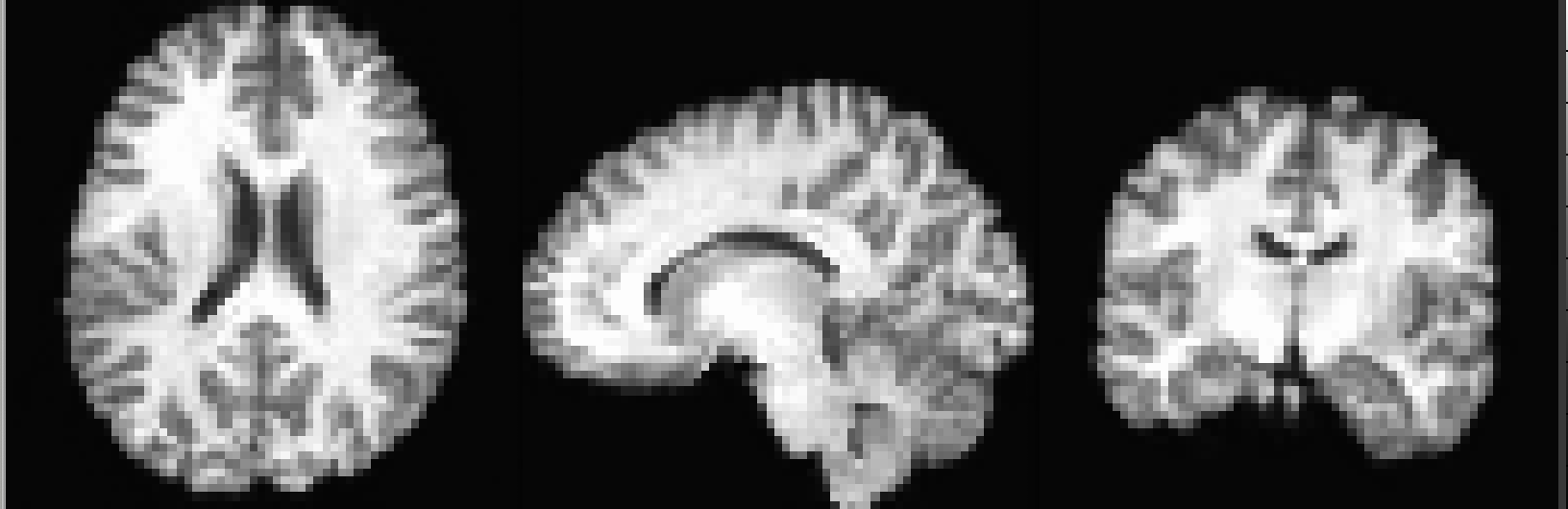}}  
\end{minipage}
\caption{Synthetic T1-weighted MR volumes with a resolution of 64 x 64 x 64 voxels (upsampled to 128 x 128 x 128 voxels) generated by our 3D progressive growing GAN.}
\label{figure2}
\end{figure*}

Figure~\ref{figure1} shows some synthetic brains with a resolution of 32 x 32 x 32 voxels generated by our 3D PGAN, while Figure~\ref{figure2} shows some synthetic brains with a resolution of 64 x 64 x 64 voxels.

\section{Conclusion}
\vspace{-0.1cm}

We have demonstrated that a 3D progressive growing GAN can be used to synthesize T1-weighted volumes of 64 x 64 x 64 voxels. The synthetic volumes can be used for training (convolutional) networks that perform classification or segmentation. Furthermore, synthetic volumes can be shared freely, as they do not belong to a specific person, and can therefore facilitate data sharing~\cite{poldrack2014making}. In future work we will synthesize volumes of higher resolution.

\vspace{-0.25cm}
\section*{Acknowledgements}

This study was supported by Swedish research council grant 2017-04889. Funding was also provided by VINNOVA Analytic Imaging Diagnostics Arena (AIDA) and the ITEA3 / VINNOVA funded project ”Intelligence based iMprovement of Personalized treatment And Clinical workflow supporT” (IMPACT).

%\vfill
%\pagebreak
% \clearpage

\clearpage
\bibliographystyle{IEEEbib}
\bibliography{refs}

\end{document}